\documentclass[letterpaper, 10 pt, conference]{ieeeconf}  

\IEEEoverridecommandlockouts                              

\usepackage{amsfonts}
\usepackage{graphicx} 
\usepackage{cite}

\usepackage{subcaption}
\usepackage{stfloats}
\usepackage{amsmath} 
\usepackage{siunitx}
\usepackage{multicol}

\makeatletter

\newcommand{\Rmnum}[1]{\expandafter\@slowromancap\romannumeral #1@}
\makeatother

\usepackage{booktabs}
\newcommand{\ra}[1]{\renewcommand{\arraystretch}{#1}}
\usepackage{mathtools}

\usepackage{cite}
\usepackage{flushend}
\usepackage{algorithm}
\usepackage{algpseudocode}
\usepackage{empheq}
\usepackage{tikz}
\usetikzlibrary{shapes.geometric, arrows}
\usepackage{makecell}
\usepackage{enumerate}
\usepackage{tcolorbox}

\algblockdefx{MRepeat}{EndRepeat}{\textbf{repeat}}{\textbf{end}}

\usepackage{etoolbox}
\makeatletter
\expandafter\patchcmd\csname\string\algorithmic\endcsname{\itemsep\z@}{\itemsep=.2ex}{}{}
\makeatother

\title{\LARGE \bf
Machine Learning-Driven Prediction of Lithium-Ion Battery Power Capability for eVTOL Aircraft
}

\author{Hao Tu{$^1$}, Yebin Wang{$^2$}, Shaoshuai Mou{$^3$}, Huazhen Fang{$^1$}
\thanks{{$^1$}Hao Tu and Huazhen Fang are with the Department of Mechanical Engineering, University of Kansas, Lawrence, KS 66045, USA.
        {\tt\small tuhao@ku.edu, fang@ku.edu}}%
\thanks{{$^2$}Yebin Wang is with the Mitsubishi Electric Research Laboratories, Cambridge, MA 02139, USA.
        {\tt\small yebinwang@ieee.org}}%
\thanks{{$^3$}Shaoshuai Mou is with the Department of Aeronautics and Astronautics, Purdue University, West Lafayette, IN 47907, USA.
        {\tt\small mous@purdue.edu}}%
}

\begin{document}

\maketitle
\thispagestyle{empty}
\pagestyle{empty}

\begin{abstract} Electric vertical take-off and landing (eVTOL) aircraft have emerged as a promising solution to transform urban transportation. They present a few  technical challenges for battery management, a prominent one of which is the prediction of the power capability of their lithium-ion battery systems. The challenge originates from the high C-rate discharging conditions required during   eVTOL flights as well as the complexity of lithium-ion  batteries' electro-thermal dynamics. This paper, for the first time, formulates a   power limit prediction problem for eVTOL which explicitly considers long prediction horizons and the possible occurrence of   emergency landings. We then harness machine learning to solve this problem in two intertwined  ways. First, we adopt a dynamic model that integrates physics with machine learning to  predict a lithium-ion battery's voltage and temperature behaviors with high accuracy. Second, while performing search for the maximum power, we leverage machine learning to predict the remaining discharge time and use the prediction to accelerate the search with fast computation. Our validation results show the effectiveness of the proposed study for eVTOL operations.

\end{abstract}
\vspace{-2mm}
\section{Introduction}

Electric vertical take-off and landing (eVTOL) aircraft hold a   promise for transforming future urban air mobility by offering a fast, convenient, and zero-carbon mode of transportation. Lithium-ion batteries (LiBs) have emerged as the technology of choice to power eVTOLs, owing to their high energy and power density. However, eVTOL flights involve aggressive power consumption profiles, necessitating the discharging of LiBs at high C-rates~\cite{Kasliwal:Nature:2019,Yang:Joule:2021,AYYASWAMY:Joule:2023}. This poses challenges for nearly all aspects of battery management. A particularly pressing and unresolved issue is how to estimate the available power of LiBs for eVTOL in flight. This issue revolves around understanding the power capability, which defines the maximum power that can be drawn from LiBs over an upcoming time horizon. Precisely identifying this capability is crucial for ensuring the safety and reliability of eVTOL operations.

The literature has presented a few studies on LiB power capability estimation, though not specifically targeting eVTOL applications. The first approach builds upon  lookup-table search~\cite{christopsen:report:2015,FARMANN:JPS:2016,Hatherall:AE:2024}. The lookup tables are constructed from data collected through offline discharging tests and map a LiB's state-of-charge (SoC) and temperature to the available power. While this approach allows easy and fast searching, it is limited in accuracy and lacks adaptability to different settings, e.g., varying horizon lengths or safety bounds, when determining power capability.

The second approach, which has recently gained growing interest, employs predictive models to achieve greater accuracy and adaptability. In this case, power capability estimation drills down to finding the current that maximizes a LiB’s power output over the prediction horizon while satisfying constraints on current, voltage, temperature, and SoC. Analytical solutions may exist when simple equivalent circuit models, such as the Rint or Thevenin model, and straightforward constraints are applied~\cite{Plett:TVT:2004,Mohan:TCST:2016,Wang:JPS:2012}. Otherwise, bisection search combined with model forward simulation can be used to find the solution, with the difficulty of search and computation varying based on the chosen models and search settings~\cite{Li:AE:2022,Dangwal:ACC:2022,Han:TTE:2022}. While many studies assume a constant maximum current magnitude over the entire horizon, the reality is that the maximum allowable current changes at different time instants due to the dynamic behaviors of LiBs. This insight has led to model predictive control (MPC)-based formulations, which maximize the power limit by optimizing the current sequence over the upcoming horizon~\cite{Zou:JPS:2018,Xavier:ACC:2020,Yang:TII:2023}. However, these MPC formulations involve running constrained optimizations that can be computationally expensive. A practical way to mitigate this is by setting up linear MPC, where applicable, to reduce computational demands~\cite{Xavier:ACC:2020}.

Despite these advances, the literature falls short of meeting the needs of eVTOL. For safety, eVTOL demands power capability estimation that is both highly accurate and computationally efficient. However, eVTOL's high discharging C-rates, reaching up to 4 to 5 C during takeoff and landing phases~\cite{Bills:SD:2023}, induce significantly more complex dynamic behaviors in LiBs, making it harder to predict their power delivery performance. While incorporating sophisticated electrochemical models in some existing methods might allow to determine the power capability, the computational demands of optimization or bisection search would be prohibitively high for real-time application. Moreover, eVTOL likely requires long prediction horizons to maintain safety, which further escalates the computational burden to an unaffordable level for existing methods.

In this paper, we present the first study on power capability estimation for eVTOL. Our specific contributions are as follows. First, we formulate a novel maximum power limit prediction problem for eVTOL. This formulation uniquely accounts for the power needed for potential emergency landings that may occur in flights, motivated by the high stakes of safety for eVTOL. This problem  necessitates  determining the highest current that can discharge the onboard LiBs over a specified time horizon without breaching safety constraints. As our second contribution, we propose to tackle this problem by: 1) identifying the remaining discharge time (RDT) under an arbitrary current, 2) comparing the RDT with the prediction horizons, and 3) using this comparison to guide a bisection search for the maximum allowable current. To facilitate this approach, we employ machine learning (ML) to develop a hybrid physics+ML dynamic model for LiBs and an RDT prediction model. Our validation results demonstrate the effectiveness of the proposed approach.

The remainder of the paper is organized as follows. Section~\ref{sec: problem formulation} proposes a novel problem formulation for LiB power prediction for eVTOL. Section~\ref{sec: power prediction} presents our ML-based LiB power prediction solution. Then, in Section~\ref{sec: Sim}, we show the simulation results based on a notional eVTOL mission profile. Finally, Section~\ref{sec: Conclusion} concludes the paper.


\begin{figure}[t!]
    \centering
    \includegraphics[width = .47\textwidth,trim={9cm 3cm 10.5cm 4cm},clip]{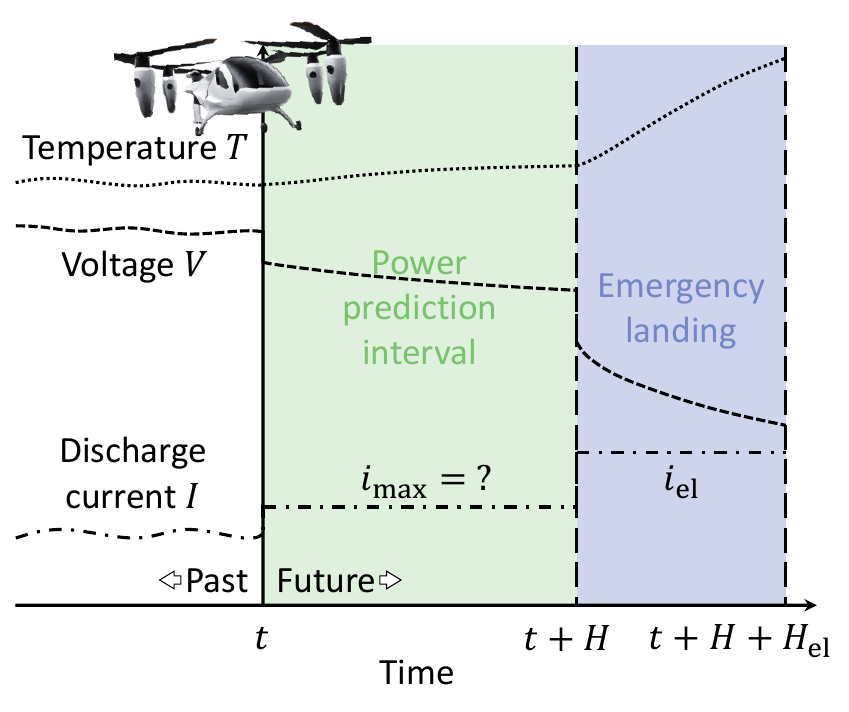}
    \caption{Proposed LiB power prediction problem formulation for eVTOL. The emergency landing is specifically considered in the predictions to account for intricate dynamics of the LiB at high C-rates.}
    \label{Fig: Powerpred}
\vspace{-5mm}
\end{figure}

\section{Problem Setup}\label{sec: problem formulation}

\begin{figure*}[t!]
    \centering
    \includegraphics[width = 0.95\textwidth,trim={.7cm 17.84cm .7cm 1cm},clip]{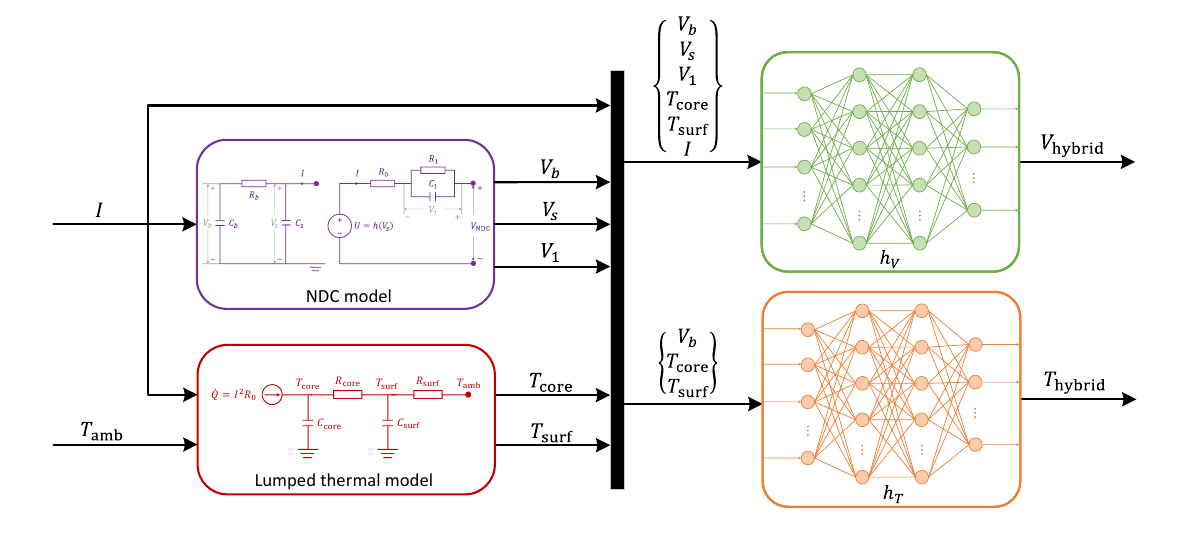}
    \caption{The NDCTNet model in~\cite{Tu:AE:2023, Tu:AE:2024}. It combines equivalent circuit models with FNNs to predict a LiB's terminal voltage and temperature behaviors. The model has been shown to have high accuracy and computational efficiency over broad C-rate ranges. }
    \label{Fig: Model}
\vspace{-5mm}
\end{figure*}

In this section, we formulate a power limit prediction problem tailored for eVTOL operation. The objective is to identify the maximum discharging power, $P_\mathrm{max}$, that a LiB can sustain over a given horizon $[t, t+H]$, where $t$ is the present time. For this horizon, $P_\mathrm{max}$ is defined as 
\begin{align*}
    P_{\mathrm{max}} = i_{\mathrm{max}}V(t+H) .
\end{align*}
where $i_{\max}$ is the maximum allowable current within the horizon. This definition implies the need to find $i_\mathrm{max}$ in order to compute $P_\mathrm{max}$. The value of $i_\mathrm{max}$ must be subject to the LiB's dynamics as well as the operational and safety constraints. Further, we propose to take into account the power requirements of emergency landing scenarios, as eVTOLs may encounter unexpected situations, e.g., system failures or adverse weather conditions, that demand reliable power availability to ensure a safe descent and landing. To address this, we introduce an additional subsequent horizon $ (t+H,t+H+H_\textrm{el}]$, where $H_\textrm{el}$ is the emergency landing horizon length, while imposing the requirement  that the LiB must be capable of discharging at $i_\textrm{el}$ within this horizon. 

Based on the above, we present the following problem formulation:
\begin{align}\label{Power-Limit-Prediction-Problem}
\begin{aligned}
i_{\mathrm{max}} = & \arg  \max   \  i \\
& \textrm{s.t.} \  \textrm{LiB dynamic model}, \\ 
& I(t + \Delta t) = i, \quad 0 \le \Delta t < H, \\
& I_{\mathrm{min}} \le i \le I_{\mathrm{max}},\\
& I(t + \Delta t) = i_\textrm{el}, \quad H \le \Delta t \le H + H_\textrm{el}, \\
& V(t + \Delta t) \ge V_{\mathrm{min}}, \quad 0 \le \Delta t \le H + H_\textrm{el}, \\
& T(t + \Delta t) \le T_{\mathrm{max}}, \quad 0 \le \Delta t \le H + H_\textrm{el} ,
\end{aligned}
\end{align}
where $V_{\min}$ and $T_{\max}$ are the voltage and temperature limits in discharging, respectively.  Fig.~\ref{Fig: Powerpred} offers a visual representation. This problem differs from its counterparts in the literature in two key aspects.  First, it incorporates two consecutive horizons: one dedicated to power limit prediction and the other for a potential emergency landing. The inclusion of this second horizon, while improving safety, adds challenges to solving the problem described in \eqref{Power-Limit-Prediction-Problem}.  Second, the problem assumes a large  $I_{\max}$  due to the high C-rate discharging characteristic of eVTOLs. Within such broad current ranges, LiBs exhibit a significant increase in variations in voltage and temperature, making the power limit more elusive to be determined.

For~\eqref{Power-Limit-Prediction-Problem}, a dynamic model is needed to describe the LiB's voltage and temperature behaviors. This model must offer high predictive accuracy across a wide range of C-rates, from low to high, while enabling fast computation in voltage and temperature prediction. The NDCTNet model proposed in~\cite{Tu:AE:2023,Tu:AE:2024} can meet these needs by integrating physics with machine learning to predict both voltage and temperature accurately. 
Fig.~\ref{Fig: Model} shows the model's structure, which couples the nonlinear double capacitor (NDC) model and a lumped thermal model, respectively, with a feedforward neural network (FNN). The NDC model is governed by
\begin{subequations} \label{NDC-SS}
\begin{align} \label{Eqn: NDC-Dynamic}
\begin{bmatrix}
\dot{V_b}(t) \\ \dot{V_s}(t) \\ \dot{V_1}(t)
\end{bmatrix}
&= A_{\mathrm{NDC}}
\begin{bmatrix}
V_b(t) \\ V_s(t) \\ V_1(t)
\end{bmatrix} 
+ B_{\mathrm{NDC}}I(t),\\ 
V_{\mathrm{NDC}}(t) &= h(V_s(t)) + V_1(t) - R_0 I(t),
\end{align}
\end{subequations}
where $V_b$, $V_s$ and $V_1$ are the voltage across $C_b$, $C_s$ and $C_1$, respectively, $I$ is the current with $I>0$ for discharge and $I<0$ for charge. Here, 
\begin{align*}
    A_\textrm{NDC} =
    \begin{bmatrix}
    \frac{-1}{C_bR_b} & \frac{1}{C_bR_b} & 0 \\
    \frac{1}{C_sR_b} & \frac{-1}{C_sR_b} & 0 \\
    0 & 0 & \frac{-1}{R_1C_1}
    \end{bmatrix}, 
    B_\textrm{NDC} =
    \begin{bmatrix}
    0\\
    \frac{-1}{C_s}\\
    \frac{-1}{C_1}
    \end{bmatrix}.
\end{align*}
The lumped thermal model is governed by
\begin{align} \label{Eqn: Thermal-Dynamic}
\begin{bmatrix}
\dot{T}_{\mathrm{core}}(t) \\ \dot{T}_{\mathrm{surf}}(t)
\end{bmatrix}
&= A_\textrm{therm}
\begin{bmatrix}
T_{\mathrm{core}}(t) \\ T_{\mathrm{surf}}(t)
\end{bmatrix} 
+ B_\textrm{therm}
\begin{bmatrix}
    I^2(t) \\
    T_{\mathrm{amb}}
\end{bmatrix},
\end{align}
where $T_{\mathrm{core/surf}}$ is the temperature at the core/surface of the cell, $T_{\mathrm{amb}}$ is the ambient temperature, and
\begin{align*}
    A_\textrm{therm} &=
    \begin{bmatrix}
    \frac{-1}{R_{\mathrm{core}} C_{\mathrm{core}}} & \frac{1}{R_{\mathrm{core}} C_{\mathrm{core}}} \\
    \frac{1}{R_{\mathrm{core}} C_{\mathrm{surf}}} & \frac{-1}{R_{\mathrm{surf}} C_{\mathrm{surf}}}+\frac{-1}{R_{\mathrm{core}} C_{\mathrm{surf}}}
    \end{bmatrix}, \\
    B_\textrm{therm} &=
    \begin{bmatrix}
    \frac{R_0}{C_{\mathrm{core}}} & 0\\
    0 & \frac{1}{R_{\mathrm{surf}} C_{\mathrm{surf}}}
    \end{bmatrix}.
\end{align*}
Both the NDC and   thermal models are linear, providing limited accuracy. They are further cascaded with two  FNNs that take the physical state as input and output the predicted voltage and temperature. These neural networks are trained on data collected from charging/discharging tests across a wide range of C-rates, thus  gaining strong predictive capability over broad current ranges. Please see~\cite{Tu:AE:2023,Tu:AE:2024} for more explanation and details. 

The NDCTNet model equations can be written compactly to take the following form:
\begin{subequations} \label{Eqn: hybrid model}
\begin{empheq}[left=\empheqlbrace]{align} \label{Eqn: hybrid dynamic model}
      &\dot{x} = f_{\mathrm{phy}}(x,I,T_{\mathrm{amb}}), \\
    &V_{\mathrm{hybrid}} = h_V(x,I), \\
    &T_{\mathrm{hybrid}} = h_T(x),
\end{empheq}
\end{subequations}
where $x = [V_b\ V_s\ V_1\ T_\mathrm{core}\ T_\mathrm{surf}]^\top$, (\ref{Eqn: hybrid dynamic model}) results from (\ref{Eqn: NDC-Dynamic}) and (\ref{Eqn: Thermal-Dynamic}), and $h_V$ and $h_T$ represent the two neural networks. Note that~\eqref{Eqn: hybrid dynamic model} is  a linear ordinary differential equation. As such, given $x(t)$, an analytical solution of $x(t+\Delta t)$ is available and expressed as
\begin{align}\label{Hybrid-model-solution}
    x(t+\Delta t) =  \phi\left(x(t), i, T_{\mathrm{amb}}, \Delta t \right).
\end{align}
The exact form of $\phi$ is shown in~\cite{Tu:AE:2024} and will find use later.

\section{ML-Driven Power Limit Prediction} \label{sec: power prediction}

In this section, we address the problem in~\eqref{Power-Limit-Prediction-Problem} to predict the power limit for eVTOL in real time. We  begin with a shortcut solution drawn from the literature and then  proceed to show the proposed approach. 

\subsection{The Shortcut Solution}

Taking cues from~\cite{Plett:TVT:2004}, we can develop a solution to~\eqref{Power-Limit-Prediction-Problem} through using bisection search of $i_{\max}$ along with model forward simulation to check the constraint satisfaction. Specifically, consider the search range of $\left[\underline{i},  \bar i \right]$. Initially, $\underline{i} = I_{\min}$, where $I_{\min}$ is the lowest possible value for $i_{\max}$, and $\bar i = I_{\max}$. Then, we run the following procedure:

\begin{tcolorbox}[left=0pt,right=0pt,top=0pt,bottom=0pt, frame empty]
\begin{algorithmic}

\MRepeat
\State $i  = ( \underline{i} +  \bar{i}) /2$
\State compute~\eqref{Hybrid-model-solution} for $\Delta t \in (0, H + H_\textrm{el}]$ 

\If{$V(t+\Delta t)\ge V_\mathrm{min}$ and $T(t + \Delta t) \le T_\mathrm{max}$ for all $\Delta t \in (0, H + H_\textrm{el}]$}
    \If{$|i - \bar i| < \epsilon$}
    \State stop, output $i_{\max} = i$
    \EndIf
    \State $\underline{i} = i$
\Else
    \State $\bar{i} = i$
\EndIf

\EndRepeat
\end{algorithmic}
\end{tcolorbox}
The above procedure provides a shortcut to treat the problem in~\eqref{Power-Limit-Prediction-Problem}. However, this shortcut comes at a cost, as it is 
 computationally expensive. The main reason is that it requires running~\eqref{Hybrid-model-solution} for all $\Delta t \in (0, H + H_\textrm{el}]$ in order to verify the constraint adherence. This causes heavy computation, as $H+H_\textrm{el}$  can be up to a few minutes.  The procedure thus is not a fitting choice for eVTOL, even though it may suit other applications.  We hence develop a new approach to achieve computationally efficient power limit prediction for eVTOL.

\subsection{The Proposed Approach}\label{sec: proposed}

The shortcut solution requires heavy computation to check whether the constraints are ensured in searching for $i_{\max}$. To facilitate computation, we introduce a  shift in perspective---rather than conducting  the tedious constraint verification, we compare the RDT with $H$ and $H_\textrm{el}$, where the RDT is subject to $i_{\max}$ and the voltage and temperature constraints. This comparison then provides a guide in searching for $i_{\max}$. With this perspective, we must predict the RDT accurately and efficiently and apply a principled search procedure. 

For a LiB, the RDT at time $t$ depends on the present condition $x(t)$, the applied current $i$, and the constraints $V_{\min}$ and $T_{\max}$. We emphasize that $V_{\min}$ and $T_{\max}$ will lead to different discharging times. We thus define the RDT as whichever of them comes first:
\begin{align}\label{eqn: VminTmax}
\Delta t_\mathrm{RDT} = \min \left\{\Delta t_\mathrm{RDT}^{V_\mathrm{min}}, \Delta t_\mathrm{RDT}^{T_\mathrm{max}}   \right\},
\end{align}
where $\Delta t_\mathrm{RDT}^{V_\mathrm{min}}$ and $\Delta t_\mathrm{RDT}^{T_\mathrm{max}}$ is the time duration that elapses before the cell reaches $V_\mathrm{min}$ and $T_\mathrm{max}$, respectively. To find out $\Delta t_\mathrm{RDT}$, we can develop a ML-based predictor in the form of
\begin{align}\label{RDT-ML-Predictor}
    \Delta t_\mathrm{RDT} = \Psi_{\mathrm{ML}}\left(x(t),i,T_{\mathrm{amb}};V_{\min}, T_{\max} \right).
\end{align}
The development of $\Psi_{\mathrm{ML}}$ will be shown later. 

Provided~\eqref{RDT-ML-Predictor}, we can perform bisection search in a new way. After initializing the search range of $\left[\underline{i},  \bar i \right]$ with $\underline{i} = I_{\min}$  and $\bar i = I_{\max}$, we implement the following:
\begin{tcolorbox}[left=0pt,right=0pt,top=0pt,bottom=0pt, frame empty]
\begin{algorithmic}
\MRepeat
\State $i  = ( \underline{i} +  \bar{i}) /2$

\State $\Delta t_\mathrm{RDT}^1 = \Psi_{\mathrm{ML}}\left(x(t), i, T_{\mathrm{amb}}; V_{\min}, T_{\max} \right)$ via~\eqref{RDT-ML-Predictor}

\If{$\Delta t_\mathrm{RDT}^1 \geq H$}

    \State $x(t+H) =  \phi\left(x(t), i, T_{\mathrm{amb}}, H \right)$ via~\eqref{Hybrid-model-solution}  

    \State  $\Delta t_\mathrm{RDT}^2 = \Psi_{\mathrm{ML}}\left(x(t+H), i_{\mathrm{el}} ,T_{\mathrm{amb}} ;V_{\min}, T_{\max} \right)$ via~\eqref{RDT-ML-Predictor}

    \If{$\Delta t_\mathrm{RDT}^2 \geq H_\textrm{el}$}
        \If{$|i-\bar i|<\epsilon$}
        \State stop, output $i_{\max} = i$
        \EndIf
        \State $\underline{i} = i$
    \Else
        \State $\bar{i} = i$

    \EndIf

    \Else
    \State $\bar{i} = i$
\EndIf
\EndRepeat
\end{algorithmic}
\end{tcolorbox}
\noindent Compared to the shortcut solution, this approach reduces the amounts of computation by substantial margins, because~\eqref{Hybrid-model-solution} is executed only once during each search iteration.

\begin{figure}[t!]
 \centering
    \includegraphics[width = 0.48\textwidth,trim={4.7cm 7.7cm 4.2cm 10.5cm},clip]{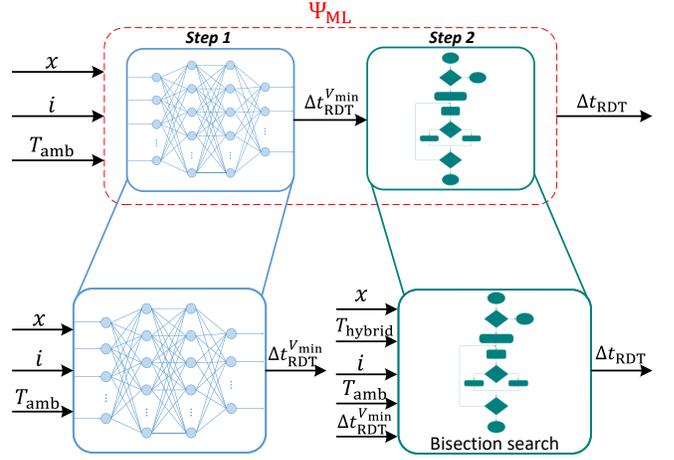}
\caption{ML-based fast RDT prediction model, denoted as $\Psi_{\textrm{ML}}$, which is used to accelerate the search for $i_\mathrm{max}$ in the proposed power prediction approach.}
\label{Fig: RDT Model}
\vspace{-5mm}
\end{figure}

In~\cite{Tu:AE:2024}, we have constructed $\Psi_{\mathrm{ML}}$ for use here and offer a review as below. As shown in Fig.~\ref{Fig: RDT Model}, $\Psi_{\mathrm{ML}}$ includes two modules in cascade. The first module is a FNN, which predicts $\Delta t_\mathrm{RDT}^{V_{\min}}$ based on $\left\{ x(t),i, T_{\mathrm{amb}} \right\}$. The neural network design is justified by theoretical analysis in~\cite{Tu:AE:2024}. After $\Delta t_\mathrm{RDT}^{V_{\min}}$ is identified, the second module seeks to determine $\Delta t_\mathrm{RDT}^{T_{\max}}$. It leverages bisection search, because the temperature changes defy data-driven prediction due to its complicated dependence on a multitude of factors, including $T_{\mathrm{amb}}$. As shown in~\cite{Tu:AE:2024}, we can train $\Psi_{\mathrm{ML}}$ using high-fidelity synthetic data generated from the NDCTNet model.

\section{Simulation Studies} \label{sec: Sim}

This section presents a validation of the proposed eVTOL power limit prediction approach. 

\subsection{Validation  Settings}
We performed experiments to collect data from a Samsung INR18650-25R cell with an NCA cathode and a graphite anode, which has a capacity of 2.5 $\textrm{Ah}$. A PEC$\textsuperscript{\textregistered}$ SBT4050 battery tester was used for charging/discharging the cell and data collection.  A thermocouple was attached to the cell's surface to measure temperature, and the ambient temperature was   maintained 
 at $T_\mathrm{amb}=25 ^\circ \textrm{C}$ throughout all tests. All data processing and power prediction simulations were performed on a host computer equipped with a 3.2 GHz Intel$\textsuperscript{\textregistered}$ i9-12900KF CPU and 128  GB RAM.

The power limit prediction problem in Section~\ref{sec: problem formulation} was set up and configured as below. The constraints are $I_\mathrm{min}=0 \ \textrm{C}$, $I_\mathrm{max}=8 \ \textrm{C}$, $V_\mathrm{min}=3\ \textrm{V}$, and $T_\mathrm{max}=50 ^\circ \textrm{C}$,  which are selected based on the manufacturer's specifications and related literature~\cite{Bandhauer:JES:2011}. Different   power prediction horizons are used, including $H=10\ \mathrm{s}$ and $3/5/7/10\ \mathrm{min}$. The minimum $10\ \mathrm{s}$ reflects the instantaneous power that can be withdrawn from the LiB. The maximum $10\ \mathrm{min}$ corresponds to the reserved cruise time required for eVTOL~\cite{Yang:Joule:2021}. The emergency landing phase is set to be $i_\textrm{el}=5 \ \textrm{C} $ and $H_\textrm{el}=105\ \mathrm{s}$ based on~\cite{Bills:SD:2023}.
    
The NDCTNet model in Section~\ref{sec: problem formulation} was identified based on the following procedures. First, we discharged the cell at low to medium C-rates (below 1 C) and collected the cell's voltage, current, and temperature data. They are used to identify the parameters of the NDC model and the lumped thermal model via least squares or maximum likelihood~\cite{Tu:ACC:2024}. Next, we performed additional discharging experiments to the cell across low to high C-rates (0$\sim$8 C) and ran the NDC and thermal model using the same current profiles. These data were used to train the $h_V$ and $h_T$ networks. We have shown in our prior study~\cite{Tu:AE:2024} that the identified NDCTNet model after training based on the above procedures has a high accuracy. The prediction step size of the NDCTNet model is 1 $\textrm{s}$. 
    
The RDT prediction model $\Psi_\textrm{ML}$ in Section~\ref{sec: power prediction} was trained as follows. We run the identified NDCTNet model under various discharging profiles to generate a synthetic dataset that contains the cell's $\Delta t_\mathrm{RDT}^{V_\mathrm{min}}$ for different state $x$ and current $i$. These data were used to train the FNN in the first module. Thereafter, $\Psi_\textrm{ML}$ is ready for making predictions. We have validated in~\cite{Tu:AE:2024} that $\Psi_\textrm{ML}$ can accurately predict the cell's RDT with good accuracy.

\subsection{Results and Discussions}

\begin{figure}[t!]
    \centering
    \includegraphics[width = .4\textwidth]{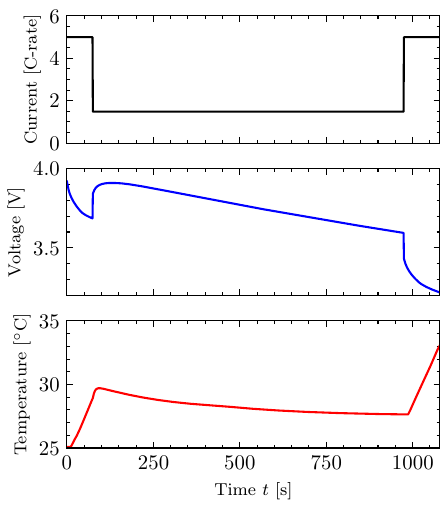}
    \caption{The current, voltage and temperature of the LiB under a notional eVTOL mission profile at $T_{\mathrm{amb}}=25$$^{\circ}\mathrm{C}$. 0$\sim$75 s: take-off; 75$\sim$975 s: cruise; 975$\sim$1080 s: landing.}
\label{Fig: eVTOL profile}
\vspace{-5mm}
\end{figure}

\begin{figure}[t!]
\centering
\begin{subfigure}{.407\textwidth}
\centering
\includegraphics[width = .98\textwidth]{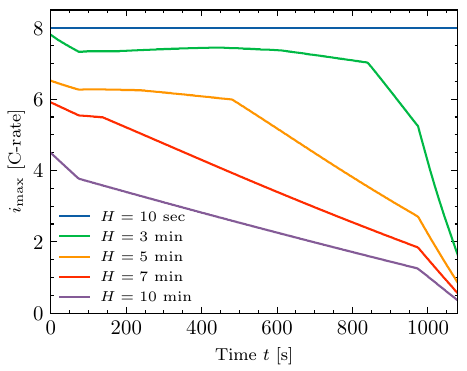}
\caption{}
\label{Fig: imax results}
\end{subfigure}

\begin{subfigure}{.42\textwidth}
\centering
\includegraphics[width = .98\textwidth]{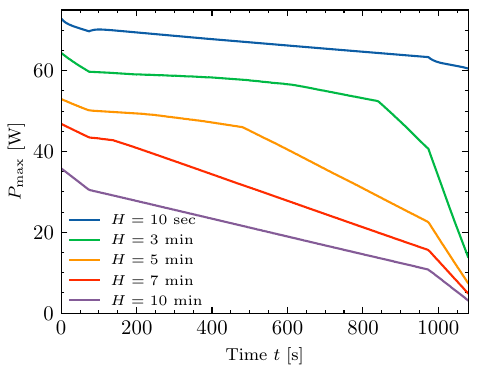}
\caption{}
\label{Fig: pmax results}
\end{subfigure}

\caption{LiB power capability prediction under a notional eVTOL profile using the proposed method. (a) The predicted $i_\textrm{max}$. (b) The resultant $P_\textrm{max}$. The $P_\textrm{max}$ reflects the maximum power load of the LiB for a cruise time of length $H$ before safely completing the landing of an eVTOL aircraft. It provides important information for the pilot to decide the landing location in an emergency.}
\vspace{-5mm}
\end{figure}

We consider a notional eVTOL mission profile as used in~\cite{Bills:SD:2023}. As shown in Fig.~\ref{Fig: eVTOL profile}, this profile is composed of 5 C, 1.48 C, and 5 C discharging phases during the take-off, cruise, and landing phases, respectively. Fig.~\ref{Fig: imax results} shows the computed $i_\textrm{max}$ at each time on the profile. We have the following observations. First, $i_\textrm{max}$ decreases during the flight. However, for $H= 10\ \textrm{s}$, $i_\textrm{max}$ is mostly equal to $I_\textrm{max}$, because the LiB would not reach $V_{\min}$ or $T_{\max}$ within in just 10 s. 
For $H= 3/5\ \textrm{min}$, one can see a slight increase after the eVTOL switch from the take-off to the cruise phase. This is because $T_\textrm{max}$, which is triggered in the take-off phase due to fast temperature rise, becomes inactive  as the LiB's temperature falls after the actual discharging current decreases from 5 C to 1.48 C in the cruise phase.  Second, the available $i_\textrm{max}$ decreases significantly as $H$ increases, which shows the necessity of considering various power prediction horizons. Fig.~\ref{Fig: pmax results} depicts the available $P_\textrm{max}$ calculated from $i_\textrm{max}$. The results show that $P_\textrm{max}$ decreases along with time and also decreases as $H$ increases. Note that this $P_\textrm{max}$ indicates the maximum power that the LiB can sustain for a horizon with length of $H$ and then completes the landing. Therefore, it shows the LiB's maximum power load for a cruise time of $H$, which is very useful for the pilot to decide on the landing location in an emergency.

\begin{figure}[t!]

\centering
\includegraphics[width = .42\textwidth]{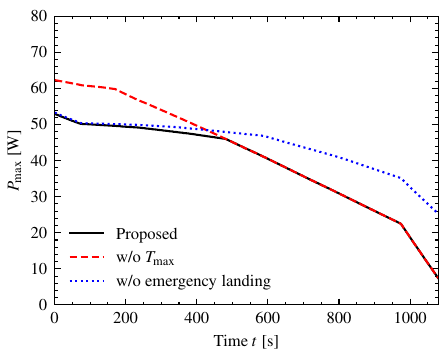}
\caption{Comparison of the LiB power prediction results under the notional eVTOL profile for different problem settings. In all cases, we set $H = 5\ \textrm{min}$. The results demonstrate the necessity of considering $T_\mathrm{max}$ and an emergency landing phase in limiting the maximum power of LiB for eVTOL.}
\label{Fig: comp results}

\end{figure}

Further, we compare our approach with two different  settings which are designed by removing either the $T_\mathrm{max}$ or the emergency landing phase from the proposed problem formulation in Section~\ref{sec: problem formulation}. The results are shown in Fig.~\ref{Fig: comp results}. We found that both of these simplifications overestimate the power performance of the LiB, which ultimately affects the safety of the eVTOL. This justifies the significance of the proposed problem formulation. 

It is also interesting to compare the proposed approach with the shortcut solution in Section~\ref{sec: power prediction}. We find that both methods yield similar results. However, as shown in Table~\ref{Table: computation}, the shortcut solution requires remarkably more computation time at each step, which becomes too high when $H$ is up to several minutes. By contrast, the proposed method exhibits a huge advantage in terms of computational speed, amenable to real-time computation as needed in ensuring the safety of eVTOL aircraft.

\vspace{-2mm}
\section{Conclusions} \label{sec: Conclusion}

\begin{table}[t!]\centering
\ra{1.2}
 \begin{tabular}{ c c c }
\toprule
 \makecell{$H$ length} & \makecell{Computation time \\ (proposed method)} & \makecell{Computation time \\ (shortcut solution)} \\
\midrule
 10 \textrm{s} & 0.1083 \textrm{s} & 2.81 \textrm{s} \\
 3 \textrm{min} & 0.7835 \textrm{s} & 14.55 \textrm{s} \\
 5 \textrm{min} & 0.5479 \textrm{s} & 27.07 \textrm{s} \\
 7 \textrm{min} & 0.6953 \textrm{s} & 40.42 \textrm{s} \\
 10 \textrm{min} & 0.3851 \textrm{s} & 60.22 \textrm{s} \\
\bottomrule
\end{tabular}
\caption{Comparison of the average computation time per time step for the proposed approach and the shortcut solution.}
\label{Table: computation}
\vspace{-5mm}
\end{table}

A key to revolutionizing future urban air mobility lies in the availability of safe, reliable, and high-performance eVTOL systems. However, eVTOL flights require discharging their onboard LiB systems at high power rates relative to the energy capacity. This, among others, makes accurate power capability estimation particularly challenging. Knowledge of power limits is crucial for safety-critical eVTOL operations, but there are few studies available. To fill in the gap, this paper introduces a power limit prediction problem tailored specifically for eVTOL applications. This problem involves long prediction horizons and considers power requirements for potential emergency landings, posing significant computational challenges to existing solutions. We develop an ML-driven approach to address this issue, based on the idea of identifying the RDT and comparing it with the prediction horizons to find out the maximum power. This approach leverages a hybrid physics-ML dynamic model for LiBs and harnesses ML techniques to enable rapid searches for the power limit. The validation results demonstrate the prediction performance and computational efficiency of the proposed approach.
\vspace{-2mm}

\bibliographystyle{ieeetr}
\bibliography{ref}

\begin{thebibliography}{10}

\bibitem{Kasliwal:Nature:2019}
A.~Kasliwal, N.~J. Furbush, J.~H. Gawron, J.~R. McBride, T.~J. Wallington, R.~D. De~Kleine, H.~C. Kim, and G.~A. Keoleian, ``Role of flying cars in sustainable mobility,'' {\em Nature Communications}, vol.~10, p.~1555, 2019.

\bibitem{Yang:Joule:2021}
X.-G. Yang, T.~Liu, S.~Ge, E.~Rountree, and C.-Y. Wang, ``Challenges and key requirements of batteries for electric vertical takeoff and landing aircraft,'' {\em Joule}, vol.~5, no.~7, pp.~1644--1659, 2021.

\bibitem{AYYASWAMY:Joule:2023}
A.~Ayyaswamy, B.~S. Vishnugopi, and P.~P. Mukherjee, ``Revealing hidden predicaments to lithium-ion battery dynamics for electric vertical take-off and landing aircraft,'' {\em Joule}, vol.~7, no.~9, pp.~2016--2034, 2023.

\bibitem{christopsen:report:2015}
J.~Christopsen, ``Battery test manual for electric vehicles,'' {\em INL/DE AC07-05ID14517 Rev}, vol.~3, pp.~1--50, 2015.

\bibitem{FARMANN:JPS:2016}
A.~Farmann and D.~U. Sauer, ``A comprehensive review of on-board state-of-available-power prediction techniques for lithium-ion batteries in electric vehicles,'' {\em Journal of Power Sources}, vol.~329, pp.~123--137, 2016.

\bibitem{Hatherall:AE:2024}
O.~Hatherall, A.~Barai, M.~F. Niri, Z.~Wang, and J.~Marco, ``Novel battery power capability assessment for improved {eVTOL} aircraft landing,'' {\em Applied Energy}, vol.~361, p.~122848, 2024.

\bibitem{Plett:TVT:2004}
G.~Plett, ``High-performance battery-pack power estimation using a dynamic cell model,'' {\em IEEE Transactions on Vehicular Technology}, vol.~53, no.~5, pp.~1586--1593, 2004.

\bibitem{Mohan:TCST:2016}
S.~Mohan, Y.~Kim, and A.~G. Stefanopoulou, ``Estimating the power capability of {L}i-ion batteries using informationally partitioned estimators,'' {\em IEEE Transactions on Control Systems Technology}, vol.~24, no.~5, pp.~1643--1654, 2016.

\bibitem{Wang:JPS:2012}
S.~Wang, M.~Verbrugge, J.~S. Wang, and P.~Liu, ``Power prediction from a battery state estimator that incorporates diffusion resistance,'' {\em Journal of Power Sources}, vol.~214, pp.~399--406, 2012.

\bibitem{Li:AE:2022}
W.~Li, Y.~Fan, F.~Ringbeck, D.~Jöst, and D.~U. Sauer, ``Unlocking electrochemical model-based online power prediction for lithium-ion batteries via gaussian process regression,'' {\em Applied Energy}, vol.~306, p.~118114, 2022.

\bibitem{Dangwal:ACC:2022}
C.~Dangwal, D.~Zhang, L.~D. Couto, P.~Gill, B.~Sebastien, W.~Zeng, and S.~J. Moura, ``Pack level state-of-power prediction for heterogeneous cells,'' in {\em 2022 American Control Conference (ACC)}, pp.~1066--1073, 2022.

\bibitem{Han:TTE:2022}
W.~Han, F.~Altaf, C.~Zou, and T.~Wik, ``State of power prediction for battery systems with parallel-connected units,'' {\em IEEE Transactions on Transportation Electrification}, vol.~8, no.~1, pp.~925--935, 2022.

\bibitem{Zou:JPS:2018}
C.~Zou, A.~Klintberg, Z.~Wei, B.~Fridholm, T.~Wik, and B.~Egardt, ``Power capability prediction for lithium-ion batteries using economic nonlinear model predictive control,'' {\em Journal of Power Sources}, vol.~396, pp.~580--589, 2018.

\bibitem{Xavier:ACC:2020}
M.~A. Xavier, A.~Kawakita~de Souza, G.~L. Plett, and M.~Scott~Trimboli, ``A low-cost {MPC}-based algorithm for battery power limit estimation,'' in {\em 2020 American Control Conference (ACC)}, pp.~1161--1166, 2020.

\bibitem{Yang:TII:2023}
Y.~Li, Z.~Wei, C.~Xie, and D.~M. Vilathgamuwa, ``Physics-based model predictive control for power capability estimation of lithium-ion batteries,'' {\em IEEE Transactions on Industrial Informatics}, vol.~19, no.~11, pp.~10763--10774, 2023.

\bibitem{Bills:SD:2023}
A.~Bills, S.~Sripad, L.~Fredericks, M.~Guttenberg, D.~Charles, E.~Frank, and V.~Viswanathan, ``A battery dataset for electric vertical takeoff and landing aircraft,'' {\em Scientific Data}, vol.~10, p.~344, 2023.

\bibitem{Tu:AE:2023}
H.~Tu, S.~Moura, Y.~Wang, and H.~Fang, ``Integrating physics-based modeling with machine learning for lithium-ion batteries,'' {\em Applied Energy}, vol.~329, p.~120289, 2023.

\bibitem{Tu:AE:2024}
H.~Tu, M.~Borah, S.~Moura, Y.~Wang, and H.~Fang, ``Remaining discharge energy prediction for lithium-ion batteries over broad current ranges: A machine learning approach,'' {\em Applied Energy}, vol.~376, p.~124086, 2024.

\bibitem{Bandhauer:JES:2011}
T.~M. Bandhauer, S.~Garimella, and T.~F. Fuller, ``A critical review of thermal issues in lithium-ion batteries,'' {\em Journal of The Electrochemical Society}, vol.~158, no.~3, p.~R1, 2011.

\bibitem{Tu:ACC:2024}
H.~Tu, X.~Lin, Y.~Wang, and H.~Fang, ``System identification for lithium-ion batteries with nonlinear coupled electro-thermal dynamics via {B}ayesian optimization,'' in {\em 2024 American Control Conference (ACC)}, pp.~1946--1951, 2024.

\end{thebibliography}

\end{document}